\documentclass[aps,graphicx,prl,twocolumn]{revtex4}
\usepackage[dvips]{graphics}
\usepackage[dvips]{graphicx}
\begin{document}
\title{An intrinsic limit to quantum coherence due to spontaneous symmetry breaking}
\author{Jasper van Wezel, Jeroen van den Brink and Jan Zaanen}
\affiliation{
Institute-Lorentz for Theoretical Physics, Universiteit Leiden,
P.O. Box 9506, 2300 RA Leiden, The Netherlands}
\date{\today}

\begin{abstract}
We investigate the influence of spontaneous symmetry breaking on the decoherence of a many-particle quantum system. This decoherence process is analyzed in an exactly solvable model system that is known to be representative of symmetry broken macroscopic systems in equilibrium. It is shown that spontaneous symmetry breaking imposes a fundamental limit to the time that a system can stay quantum coherent. This universal timescale is $t_{spon} \simeq 2\pi N \hbar / (k_B T)$, given in terms of the number of microscopic degrees of freedom $N$, temperature $T$, and the constants of Planck ($\hbar$) and Boltzmann ($k_B$).
\end{abstract}
\maketitle

{\it Introduction.}
The relation between quantum physics at microscopic scales and the classical behavior of macroscopic bodies has been a puzzle in physics since the days of Einstein and Bohr. This subject has revived in recent years both due to experimental progress, making it possible to study this problem empirically, and because of its possible implications for the use of quantum physics as a computational resource~\cite{Bennett00}. This `micro-macro' connection actually has two sides. Under equilibrium conditions it is well understood in terms of the mechanism of spontaneous symmetry breaking. But in the dynamical realms its precise nature is still far from clear. The question is, can spontaneous symmetry breaking play a role in a {\it dynamical} reduction of quantum physics to classical behavior? This is a highly non trivial question as spontaneous symmetry breaking is intrinsically associated with the difficult problem of many particle quantum physics. Here, we will analyze a tractable model system which is known to be representative of macroscopic systems in equilibrium, to find the surprising outcome that {\it  spontaneous symmetry breaking imposes a fundamental limit to the time that a system can stay quantum coherent}~\cite{Caldeira81,Chuang95}. This universal timescale turns out to be $t_{spon} \simeq 2\pi N \hbar / (k_B T)$.

This result is surprising in the following sense. Consider a macroscopic body at room temperature; $\hbar / (k_B T) \simeq 10^{-14}$ seconds which is quite a short time. However, multiplying it with Avogadro's number $N \simeq 10^{24}$, $t_{spon}$ becomes $ \simeq 10^{10}$ seconds, a couple of centuries. Given all other sources of decoherence for such a large macroscopic body, this is surely not a relevant timescale. However, quantum systems of contemporary interest are typically much smaller. Let us for instance consider a flux state qubit with a squid the size of one cubic micron and a temperature of the order of one Kelvin~\cite{Chiorescu03}. The time $t_{spon}$ then turns out to be of order of seconds, a coherence timescale which might well be reached in the near future. The counterintuitive feature of this intrinsic decoherence mechanism linked to equilibrium classicality is that it starts to matter when systems become small.

{\it Spontaneous symmetry breaking.}
In main-stream quantum measurement theory, the nature of the classical machine executing the measurement is treated rather casually -- it is just asserted to exist, according to daily observations. However, eventually this machine is also subjected to the laws of quantum physics. After all, it is made of microscopic stuff similar to the small quantum system on which the machine acts. The description of this machine typically involves $10^{24}$ strongly interacting quantum particles and this is not an easy problem. In fact, the very existence of the machine seems to violate the basic laws of quantum physics. The most fundamental difference between quantum- and classical physics lies in the role of symmetry. Dealing with an exact quantum mechanical eigenstate, all configurations equivalent by symmetry should have the same status in principle, while in a classical state one of them is singled out. For example, given that space is translationally invariant, the measurement machine should be in an eigenstate of total momentum, being spread out with equal probability over all of space. In the classical limit however it takes a definite locus. The explanation of this `spontaneous symmetry breaking' in terms of the singular nature of the thermodynamic limit is one of the central achievements of quantum condensed matter physics~\cite{Anderson72}. One imagines a symmetry breaking `order parameter field' $h$ (e.g., a potential singling out a specific position in space). Upon sending $h$ to zero before taking the thermodynamic limit ($N \rightarrow \infty$) one finds the exact quantum groundstate respecting the symmetry. However, taking the opposite order of limits one finds that the classical state becomes fact. 

What does all of this have to do with the dynamical phenomenon of decoherence? Decoherence refers to the fact that the quantum information encoded in some microscopic state entangles in the course of its time evolution with environmental degrees of freedom. Since this information cannot be recovered `for any practical purpose', one should trace out the environment from the density matrix with the effect that the reduced density matrix will reveal a mixed state. The crucial point is that spontaneous symmetry breaking is intrinsically linked to the presence of a spectrum of `environmental states'. In a rigorous fashion, the quantum information carried by these states cannot be retrieved when the body is macroscopic. This so-called `thin spectrum' is so sparse that it even ceases to influence the partition function~\cite{Kaplan90}. The question we wish to address in the remainder is, to what extent this {\em thin spectrum can be a source of decoherence, intrinsically associated with the fact that quantum measurements need classical measurement machines}.

Given that spontaneous symmetry breaking involves the a-priori untractable problem of a near-infinity of interacting quantum degrees of freedom, this question cannot be answered in full generality. However, some time ago it was discovered that the mechanism of spontaneous symmetry breaking reveals itself in representative form in a simple, integrable model. This model is the Lieb-Mattis long-ranged quantum Heisenberg antiferromagnet~\cite{Lieb62}, given by the Hamiltonian
\begin{eqnarray}
H_{LM} = \frac{2|J|}{N} {\bf S}_A \cdot {\bf S}_B - h (S_A^z - S_B^z),
\end{eqnarray}
It is defined for a bipartite lattice with $A$ and $B$ sublattices, where ${\bf S}_{A/B}$ is the total spin on the $A/B$ sublattice with $z$-projection $S_{A/B}^z$, and $h$ is the symmetry breaking field, in this case a staggered magnetic field acting on the staggered magnetization $M^z = S_A^z - S_B^z$. The particularity of the Lieb-Mattis Hamiltonian is that every spin on sublattice $A$ is interacting with {\it all} spins on sublattice $B$ and vice versa, with interaction strength $2|J|/N$ (which depends on the total number of sites $N$ so that the system is extensive). This very simple Hamiltonian accurately describes the thin spectrum encountered in more complicated Hamiltonians, like the nearest neighbor Heisenberg antiferromagnet, the BCS superconductor, or the harmonic crystal~\cite{Bernu92, Anderson58}. Moreover, in this Hamiltonian the singular nature of the thermodynamic limit can be explicitly demonstrated~\cite{Kaplan90, Kaiser89}. We will therefore use the Lieb-Mattis magnet as a model for a measurement machine.

{\it Measurement scheme.}
Our scheme for quantum measurement using this Lieb-Mattis magnet explicitly keeps track of the particular role of the thin spectrum. We start out preparing the Lieb-Mattis machine built from $N$ spins at time $t < t_0$ in the symmetry broken N\'eel ground state ($\langle M^z \rangle \neq 0$) of $H_{LM}$. The microscopic quantum system to be measured is isolated at $t<t_0$ and consists of two qubits (qubits $a$ and $b$, each with two $S=1/2$ states) in a maximally entangled singlet state, $|qubit \rangle = \frac{1}{\sqrt{2}}\left[ | \uparrow_a \downarrow_b \rangle - | \downarrow_a \uparrow_b \rangle \right]$. At time $t=t_0$ we instantaneously include qubit $a$ ($b$) in the Lieb-Mattis (infinite range) interactions of the spins on the $A$ $(B)$ sublattice of the Lieb-Mattis machine. We then follow the exact time evolution of the coupled $N+2$ particle system at $t > t_0$:
\begin{eqnarray}
H_{t<t_0} &=& H_{LM}+ {\bf S}_a \cdot {\bf S}_b \nonumber \\
H_{t>t_0} &=& \frac{2|J|}{N+2} {\bf S}_{A+a} \cdot {\bf S}_{B+b} - h (S_{A+a}^z - S_{B+b}^z),
\end{eqnarray}
where ${\bf S}_{A+a}$ is ${\bf S}_A + {\bf S}_a$, and ${\bf S}_{B+b}$ is ${\bf S}_B + {\bf S}_b$.

To obtain further insight in how this quantum measurement works, let us first see what would happen if the measurement machine would be semi-classical, i.e. described in terms of a spin wave expansion. This starts with assuming a maximally polarized staggered magnetization $\langle M^z \rangle$ for the Lieb-Mattis measurement machine. By linearizing the equations of motion one then obtains the spin waves that are characterized by a 'plasmon' gap due to the long range nature of the interactions. Stronger, because of the infinite range of the interactions their spectrum is dispersionless and it is easily demonstrated that in fact the spin waves do not give rise to perturbative quantum corrections to the staggered magnetization --from this perspective, the classical N\'eel state appears to be an exact eigenstate. It is now immediately clear what happens at times $t > t_0$. At $t<t_0$ the system was prepared in a product state of the spin singlet qubit and the $N$-spin N\'eel ground state of the Lieb-Mattis antiferromagnet: $|\psi_{t<t_0} \rangle = |0\rangle_{N} \times |qubit\rangle$. When at $t=t_0$ the interaction between the micro and macro system is switched on, the $N+2$ spin system can either be in its N\'eel ground state $|0 \rangle_{N+2}$ or in an excited state where both spins $a$ and $b$ are misaligned relative to the magnetization on the respective sublattices with which they interact (Fig. 1). This state corresponds to a two magnon excited state and since the magnons do not propagate, this excited state $|2 \rangle_{N+2}$ also appears to be an exact eigen state. Hence, the semi-classical wavefunction is simply $|\psi^{\rm sc}_{t=t_0} \rangle = \left[ |0 \rangle_{N+2} - |2\rangle_{N+2} \right] / \sqrt{2}$ and the time evolution at $t > t_0$ is characterized by a coherent oscillation between the two states. Since the state $|2\rangle$ is distinguishable from the ground state $|0\rangle$, it is in principle measurable by slowly switching on interactions with other environmental degrees of freedom, and eventually the wavefunction will collapse. The outcome of this experiment would be the usual Rabi oscillations, with a frequency that is proportional to $E_2-E_0$, the energy difference between the two states. Thus in a semi-classical description there is  no loss of quantum coherence.

One recognizes in the above the typical way that canonical measurement machines are conceptualized in quantum measurement theory. The surprise is now that even for this (in a sense, extremely `classical') Lieb-Mattis measurement machine the preceding semi-classical analysis is only exact when {\em the machine is infinitely large}! The construction turns out to be subtly flawed when $N$ is finite and $T>0$. The culprit is the thin spectrum which is completely disregarded in the semi-classical analysis. To reveal the decohering effect of the thin spectrum, the Lieb-Mattis model should be solved exactly. This can in fact easily be done by first introducing the operator of {\em total} spin ${\bf S}={\bf S}_A+{\bf S}_B$. Taking $h=0$, the Hamiltonian can then be written as $(J/N) ( S^2 - S^2_A -S^2_B)$ and accordingly the eigenstates are $|S_A, S_B, S, M \rangle$ where $S,M$ denote total spin and its z-axis projection, while $S_A$ and $S_B$ refer to the total sublattice spin quantum numbers. $S_A$ and $S_B$ are maximally polarized in the ground state. Lowering $S_A$ or $S_B$ corresponds to exciting a magnon carrying an energy $J$. One sees immediately that the {\em true} ground state of the system is an overall $S=0$ {\em singlet}, i.e. a state characterized by $\langle M^z \rangle = 0$. One also infers the presence of a tower of {\em total} $S$ states characterized by an energy scale $E_{thin}=J/N$, and this is the thin spectrum. For a finite staggered magnetic field $h$, the situation changes drastically; $h$ couples the states in the thin spectrum and it is easy to show that the ground state becomes a wave packet of thin spectrum states and in this case $E_{thin}=\sqrt{Jh}$. This groundstate does carry a finite staggered magnetization: it is the antiferromagnetic N\'eel state. One can now straightforwardly demonstrate the singular nature of the thermodynamic limit~\cite{Kaplan90,Kaiser89}. By sending first $h \rightarrow 0$ and then $N \rightarrow \infty$ one obtains the exact total singlet groundstate, respecting the spin rotational symmetry. Upon taking the opposite order one finds the fully polarized N\'eel antiferromagnet of the semi-classical expansion.

{\it Exact time evolution.}
Let us now reconsider our quantum measurement, taking full account of the thin spectrum states (Fig.~2). For $t<t_0$ the Lieb-Mattis machine is described by the following thermal density matrix, assuming that ${\rm k_B T} \ll J$ so that magnon excitations can be neglected, 
\begin{eqnarray}
\rho_{t < t_0} = \frac{1}{Z} \sum_{n=0}^{N-1} e^{-\frac{E^n_0}{\rm k_B T } } |0,n \rangle \times |qubit \rangle \langle 0,n| \times \langle qubit|, 
\end{eqnarray}
where $Z$ is the partition function, the thin spectrum states are labeled by $n$ and have an energy $E_0^n$. Switching on the Lieb-Mattis interaction between the qubits and the machine's sublattices at $t = 0$ we find that the density matrix at $t > t_0$ becomes, 
\begin{eqnarray}
\rho_{t>t_0} &=& U \rho_{t=t_0} U^{\dagger} \nonumber \\
&=& \frac{1}{2Z} \sum_{n=0}^{N-1} \left. e^{-\frac{E^n_0}{\rm k_B T} }\right. \left[ |0,n\rangle \langle 0,n| + |2,n\rangle\langle 2,n| \right. \nonumber \\
&+& e^{-i(E^n_2-E^n_0)(t-t_0)/\hbar } \left. \left( |0,n\rangle\langle 2,n| +h.c.\right) \right], 
\end{eqnarray}
where $U$ is the exact time evolution operator and the states now describe the $(N+2)$-particle Lieb-Mattis model. Given their unobservable nature~\cite{Kaplan90}, we trace over the thin spectrum states in this density matrix. The off-diagonal matrix elements of this reduced density matrix are now
\begin{eqnarray}
\rho^{OD}_{t>t_0}=\frac{e^{-2iJ(t-t_0)/\hbar}}{2Z} \sum_{n=0}^{N-1} e^{-\frac{E^n_0}{\rm k_B T} } 
e^{-i(E^n_2-E^n_0-2J)(t-t_0)/\hbar }, 
\end{eqnarray}
where the phase factor associated with the two-magnon state is taken out of the summation. The absolute value $|\rho^{OD}_t|$ is the measure for the time dependent entanglement between states $|0\rangle$ and $|2\rangle$. It can be evaluated exactly for any given $N$ and the result is shown in Fig.~3. The vanishing of this matrix element in the course of the time evolution signals decoherence and we find that this is associated with a characteristic timescale of a remarkably universal nature: under the physical conditions that $E_{thin} \ll {\rm k_B T} \ll J$ and $J/N < hN$ we find that the decoherence time due to spontaneous symmetry breaking becomes completely independent of the energy scales characterizing the system: $t_{spon} = { 2 \pi N \hbar}/{ \rm k_B T}$, the result we announced in the beginning.

The fact that the reduced density matrix at $t>t_0$ describes a mixed state, while at $t<t_0$ the system was in a pure state, could lead to the conclusion that the present mechanism for decoherence is irreversible. But irreversibility is at odds with unitary time evolution~\cite{Adler02}. We actually do find that after a certain time $t_{rec}$ the system returns to a pure state again, with exactly the same reduced density matrix it started with at $t=t_0$. Thus the decoherence is in fact reversible, see fig 3. This recurrence time depends on the energy scales of the Lieb-Mattis measurement machine in a quite remarkable way: $t_{rec}/ t_{spon}= {\rm k_B T} / E_{thin}$. Under the physical condition that the typical level splittings in the thin spectrum are very small compared to temperature, the recurrence takes infinitely long so that for all practical purposes the thin spectrum acts as a truly dissipative bath turning quantum information into an increase of classical entropy. 

{\it Origin of decoherence.}
Given that decoherence via the thin spectrum requires temperature to be finite, it is tempting to associate $t_{spon}$ with the thermal fluctuations of the order parameter in the finite system, as described by spin wave theory. However, this is {\em not} the case because these thermal fluctuations invoke the thermal excitation of the magnon states. These are exponentially suppressed by Boltzmann factors $e^{-J/(k_B T)}$, which depend on the energy scale $J$ of the individual interactions. The origin of $t_{spon}$ is more subtle: it is due to the hidden thin spectrum that reflects the zero point fluctuations of the order parameter as a whole. This thin spectrum does not carry any thermodynamic weight, and turns into a heat bath destroying quantum information if temperature is finite.

It is remarkable that the coherence time is such a universal timescale, independent of the detailed form of the thin spectrum --which, after all, is determined by the parameters $J$ and $h$ in the Lieb-Mattis Hamiltonian. Physically one can think of this universal timescale as arising from two separate ingredients. First, the energy of a thin spectrum state $|n \rangle$ changes when magnons appear.  The change is of the order of $n E_{thin}/N$, where $E_{thin}$ is the characteristic level spacing of the thin spectrum that we happen to be considering. The fact that each thin state shifts its energy somewhat at $t>t_0$ leads to a phase shift of each thin state and in general these phases interfere destructively, leading to dephasing and decoherence. The larger $n E_{thin}/N$, the faster this dynamics. But in order for this dephasing to occur, it is necessary for a finite number of thin states to actually participate in the dynamics of decoherence. Since temperature is finite (but always small compared to the magnon energy) a finite part of the thin spectrum is available for the dynamics. Thin spectrum states with an excitation energy higher than $k_B T$ are suppressed exponentially due to their Boltzmann weights. The maximum number of thin states that do contribute is roughly determined by the condition that $n^{max} \sim k_B T/E_{thin}$. Putting the ingredients together, we find that the highest energy scale that is available to the system to decohere is approximately $ \frac{k_B T}{E_{thin}} \frac{E_{thin}}{N}$. All together, the thin spectrum drops out of the equations. The fastest time scale at which the dynamics take place is given by the inverse of this energy scale, converted into time: one finds the decoherence time $t_{spon} \sim \frac{2 \pi \hbar N}{k_B T}$.

{\it Conclusions.}
To what extent is the Lieb-Mattis machine representative of a general classical measurement machine displaying a broken continuous symmetry? In fact the Lieb Mattis machine is the best case scenario for the kind of measurement machine envisaged in main stream quantum measurement theory, as its behavior is extremely close to semi-classical due to the presence of the infinite range interactions. Machines characterized by short range interactions carry massless Goldstone modes and these will surely act as an additional heat bath limiting the coherence time. It is of course not an accident that the most 'silent' systems are qubits based on superconducting circuitry, which have a massive Goldstone spectrum in common with the Lieb-Mattis system. We have demonstrated here that even under these most favorable circumstances quantum coherence eventually has to come to an end, because of the unavoidable condition that even the most classical measurement machines are subtly influenced by their quantum origin. These effects become noticeable in the mesoscopic realms and we present it as a challenge to the experimental community to measure $t_{spon}$.

\pagebreak

\begin{figure}[ht]
\includegraphics[width=0.8\columnwidth]{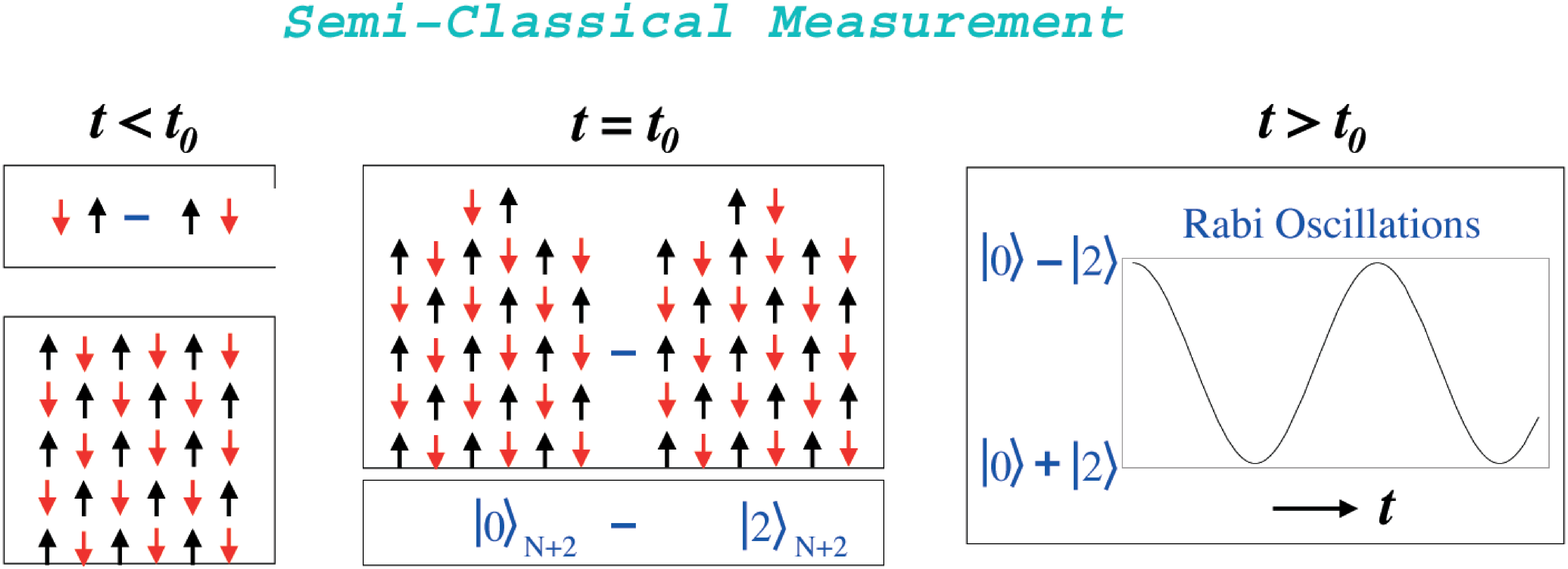}
\caption{Semi-classical time evolution of a two spin qubit that at $t=t_0$ starts interacting with a Lieb-Mattis measurement machine. Quantum coherence is preserved at all times.} 
\end{figure}

\begin{figure}[ht]
\includegraphics[width=0.4\columnwidth]{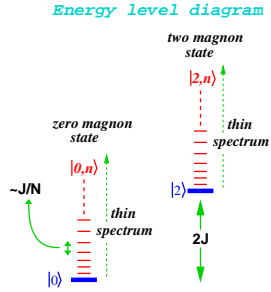}
\caption{Energy level scheme with the zero and two magnon states, each with its tower of thin spectrum states. The level spacing in the thin spectrum is $E_{thin}$, magnons live on an energy scale $J$.} 
\end{figure}

\begin{figure}[ht]
\includegraphics[width=0.5\columnwidth]{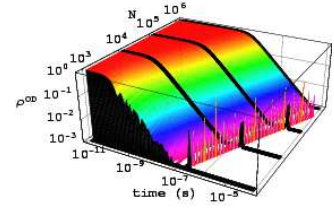}
\includegraphics[width=0.5\columnwidth]{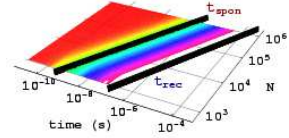}
\caption{The time dependence of the entanglement between states $|0\rangle$ and $|2\rangle$, $|\rho^{OD}|$, for different numbers of spins $N$ at T=10 Kelvin and $t_{rec}/t_{spon}=10^3$. In the bottom figure the decoherence time due to spontaneous symmetry breaking $t_{\rm spon}$ and the recurrence time $t_{\rm rec}$ are indicated.} 
\end{figure}

\end{document}